\documentclass[prb,twoside,twocolumn,amsmath,amssymb]{revtex4}

\newif\ifrevtex
\ifx\pacs\undefined
\revtexfalse 
\else
\revtextrue
\fi

\ifrevtex
\let\citeonline\onlinecite
\else
\usepackage{cite} 
\usepackage{overcite}
\usepackage{geometry}
\let\email\footnote
\let\homepage\footnote
\let\affiliation\footnote
\fi

\usepackage{amsmath} 
\usepackage{bm}
\usepackage{verbatim} 
\usepackage{tabularx} 

\usepackage[tight]{subfigure}
\usepackage[import]{xy}
\usepackage{graphicx}
\DeclareGraphicsExtensions{.pdf,.eps}

\newcommand{\myhref}[1]{\href{#1}{#1}}
\newcommand{\mailto}[1]{\href{mailto:#1}{#1}}

\renewcommand{\v}[1]{\ensuremath{\mathbf{#1}}} 

\newcommand{\subfigfourth}[1]{\includegraphics[width=3.765cm]{#1}}
\newcommand{\subfigfourthnlabel}[2]{\leavevmode\begin{xy}\xyimport(100,100){\subfigfourth{#1}},(20,98)*!LU\txt{#2}\end{xy}}
\newcommand{\fourfigs}[2]{%
\subfigfourth{#1}%
\subfigure{\label{subfig:#1}\subfigfourthnlabel{#1ASSFx}{(\alph{subfigure}) $\gamma=#2$}}%
\subfigfourth{#1ASSFy}%
\subfigfourth{#1ASSFz}%
\\%
}

\begin{document}

\title{Structural changes in block copolymer solution under shear flow as determined by nonequilibrium molecular dynamics}

\ifrevtex
\author{Igor Rychkov}
 \email{rych@chem.scphys.kyoto-u.ac.jp}
\author{Kenichi Yoshikawa}
 \homepage{http://www.chem.scphys.kyoto-u.ac.jp}
\affiliation{%
Department of Physics, Graduate School of Science,\\ Kyoto University, Kyoto 606-8502, Japan.}%
\else
\author{Igor Rychkov\footnote{\mailto{rych@chem.scphys.kyoto-u.ac.jp}} and Kenichi Yoshikawa\footnote{\myhref{http://www.chem.scphys.kyoto-u.ac.jp}}}
\begin{center}
\textsl{Graduate School of Science, Department of Physics, Kyoto University, Kyoto 606-8502, Japan}
\end{center}
\fi
\date{\today}
\begin{abstract}
A nonequilibrium molecular dynamics computer simulation on microsegregated solutions of symmetrical diblock copolymers is reported. As the polymer concentration increases, the system undergoes phase transitions in the following order: body centered cubic (BCC) micelles, hexagonal (HEX) cylinders, gyroid (GYR) bicontinuous networks, and lamellae (L), which are the same morphologies that have been reported for block copolymer melts. Structural classification is based on the patterns of the anisotropic static structure factor and characteristic 3-dimensional images. The systems in the BCC micellar ($\rho\sigma^{3}=0.3$) and HEX cylindrical ($\rho\sigma^{3}=0.4$) phases were then subjected to a steady planar shear flow. In weak shear flow, the segregated domains in both systems tend to rearrange into sliding parallel  close-packed layers with their normal in the direction of the shear gradient. At higher shear rates both systems adopt a perpendicular lamellar structure with the normal along the neutral direction. A further increase in the shear rate results in a decrease in lamellar spacing without any further structural transitions. Two critical shear rate values that correspond to the demarcation of different structural behaviors were found.
\end{abstract} 
\maketitle

\section{Introduction}
Block copolymers are macromolecules composed of sequences, or blocks, of chemically distinct repeat units. Due to the low entropy of macromolecules, even a small distinction between the units causes, below the order-disorder transition (ODT) temperature, a phase separation. Because of the chemical connectivity of the blocks, block copolymers cannot separate into macroscopic phases. Instead, spatially periodic nanostructures are formed. The structure and its symmetry, or morphology, is determined by the volume fraction of blocks, or ultimately by the copolymer composition.

We shall consider the simplest type of block copolymers: linear $AB$ diblock copolymers. In this case, the copolymer composition is characterized by the ratio $f$ of the number of monomeric units in the block of species $A$, $N_{A}$, to the overall degree of polymerization, $N = N_{A} + N_{B}$: $f = N_{A} / N$. We would like to characterize the possible equilibrium morphologies of block copolymer systems by referring to block copolymer melts, which have been fairly well studied both experimentally and theoretically.\cite{BF99, Ham99} Sorted in ascending $f$, these morphologies are cubic (BCC or FCC) packing of spherical micelles (S), hexagonally packed cylinders (C), bicontinuous cubic (``gyroid") phase (G) , and lamellae (L), with the last corresponding to the symmetrical case of $f = 1/2$.


When diblock copolymers are dissolved in a selective solvent, i.e. a solvent favorable for one component (e.g.,  $A$) and poor for the other ($B$), the morphologies also depend on the concentration of the block copolymers. The full phase diagram has been successfully obtained experimentally \cite{L02} and confirmed by self-consistent mean-field (SCMF) theory.\cite{L98} Interestingly, the morphological changes that occur by varying the concentration of diblock copolymers in a selective solution are essentially the same as those that occur in melt by varying the composition. Above the critical micelle concentration (CMC), symmetrical diblock copolymers in a selective solvent form spherical micelles which, with an increase in concentration, arrange into cubic (BCC or FCC) micellar crystals, and then successively transform to hexagonally ordered cylinders, bicontinuous networks, and, finally, lamellae (Disordered micelles $\to$ S $\to$ C $\to$ G $\to$ L). The similarity of the two scenarios can be understood by the following simple consideration. The addition of a selective solvent (or increasing the selectivity of the solvent) to the melt of block copolymers corresponds to an increase in the volume fraction of the selected block, and thus the phase behavior of the solution as a function of the concentration can be mapped onto the melt with increasingly asymmetric copolymer compositions.


Effects of shear on block copolymer systems include shear-induced ordering, domain reodering, and disordering. In the isotropic-lamellar transition, by suppressing nonlinear fluctuations, shear flow raises the transition temperature; thus, in a certain temperature region the lamellar phase can be induced by applying shear. The lamellar ordering occurs with wave vector normal to both shear and velocity gradient. \cite{CatesMilner89,Bates93}

Reordering of block copolymer nanostructures to new low-frictional phases occurs when the flow becomes faster than the relaxation rate of the system, for a review of recent experimental reports see Ref.~[\citeonline{H00,H01}]. The lamellar phase under a weak planar shear flow (low shear rate or low shearing frequency) adopts an orientation parallel to the shear plane, with the  lamellar normal along the flow gradient direction, whereas at higher shear rates, the lamellae flip to the perpendicular orientation, with their normal along the flow vorticity (neutral) direction.\cite{F94} A further increase in the shear rate results in a reduction in lamellar spacing. When the hexagonal cylindrical phase is subjected to  shear flow, the cylinders are perfectly oriented along the flow direction, with the hexagonal plane \{10\} either parallel, at low shear rates, or perpendicular, at higher shear rates and near ODT, to the flow. It has also been shown that spatial fluctuations of the cylindrical domains, if characterized by a lifetime greater than the inverse shear rate, lead to break-up, or disordering, which has in fact been observed at higher shear rates. Cubic micellar phases under shear flow change to more complicated symmetries, such as twinned BCC and FCC lattices and hexagonal close-packed (HCP) layers, with their densest planes (e.g. \{110\} for BCC and \{111\} for FCC) slipping in parallel to the shear plane. At higher shear rates, a loss of long-range order due to the shear-induced micelle melting has been reported.\cite{H00,H01}


Although the reorganization and reorientation of domains in microphase-segregated block copolymer systems have been reported, there have been few studies on the breakup of  domains. Since domains remain intact and due to the aforementioned mapping, previous results regarding block copolymers under shear can be reviewed without separating the results related to melts or solvents. We sought to study block copolymers in a  selective solvent under a high enough shear so that we could observe domain breakup and the corresponding new phase transitions. 

A mesoscopic simulation technique based on the dynamic mean-field density functional theory has been recently employed to investigate lamellar orientation in block
copolymers subjected to shear and was able to capture parallel and perpendicular lamellar states at low and higher shear rates.\cite{Fraaije02}

In this work, we use the non-equilibrium molecular dynamics (NEMD)\cite{EM90,Hey98} computer simulation method to model block copolymers under shear flow conditions and investigate structural changes in the systems. The data regarding rheological and other properties obtained for the same systems will be published elsewhere.


\section{Model and Method}
\subsection{Model potentials}
Using the bead-spring model, we studied systems of $n$ symmetrical diblock copolymer chains with $N = N_{A} + N_{B}$ beads per chain and a composition $f \equiv N_{A} / N = 0.5$. The selectivity of the solvent is modeled using different parameters for the  Lennard-Jones (LJ) potential. The bead pairs $A-A$ and $A-B$ interact via purely repulsive, truncated at the minimum and shifted LJ potential (Weeks-Chandler-Andersen  potential):
\begin{equation}
U_{\rm{LJ}}^{\rm rep}  = \left\{
\begin{aligned}
 &4\epsilon \left[ \left( \frac{\sigma }{r} \right)^{12}  - \left( \frac{\sigma }{r} \right)^{6}  + \frac{1}{4} \right], & r \leq 2^{1/6} \sigma\\ 
 &0, & r \geq 2^{1/6} \sigma
\end{aligned} \right.
\label{eq:LJrep}
\end{equation}
Since $\epsilon$ can be used to set the energy scale and $\sigma$ can be used for the length scale, we will henceforth use so-called Lennard-Jones reduced units where $\epsilon = \sigma = 1$ and also set the mass $m$ of the particles to unity, so that time is measured in units of $(\sigma^{2} m / \epsilon)^{1/2}$.
$B-B$ interaction includes long-ranged attraction and is modeled by the following form of LJ potential, modified so as to satisfy the continuity and smoothness requirements at the truncation point:
\begin{equation}
U_{\text{LJ}} = \left\{
\begin{aligned}
&\alpha\phi \left[ \frac{1}{r^{12}}  - \frac{1}{r^{6}} + \right. \\& \left.\frac{r^{2}}{r_{c}^{2}}\left( \frac{6}{{r_{c}}^{12}} - \frac{3}{{r_{c}}^{6}}\right)  - \frac{7}{r_{c}^{12}} + \frac{4}{r_{c}^{6}}\right], & r \leq r_{c}\\ 
&0, & r \geq r_{c} 
\end{aligned}
\right.
\label{eq:LJ}
\end{equation}
The radius of interaction is chosen to be sufficiently small, $r_{c} = 2$, to avoid possible ``freezing" of multi-particle aggregates and to increase the speed of calculation.  The potential has a minimum at $r\approx2^{1/6}$;  $\alpha=4.913708$ gives $\phi$ as the well depth. We take the temperature and well depth as $T = 1$ and $\phi/T = 1.2$, which implies that beads $A$ and $B$ are in favorable and unfavorable solvent conditions, respectively. 

For any two adjacent beads along the chain, in addition to the excluded volume LJ potential, Eq.~(\ref{eq:LJrep}), the attractive anharmonic FENE ("finitely extensible non-linear elastic") spring potential is also used with spring constant $k=30$ and maximum extension $R_{0}=1.5$:
\begin{equation}
U_{\text{FENE}} =
\left\{
\begin{aligned}
 &-\frac{1}{2} k R_{0}^{2} \ln \left[1 - \left(\frac{r}{R_{0}}\right)^{2} \right], & r \leq R_{0}\\ 
 &\infty, & r \geq R_{0}
\end{aligned}
\right.
\label{eq:FENE}
\end{equation}

\subsection{Equations of motion}
\begin{figure}\centering
\leavevmode\begin{xy}\xyimport(100,100){\includegraphics[width=4.8cm]{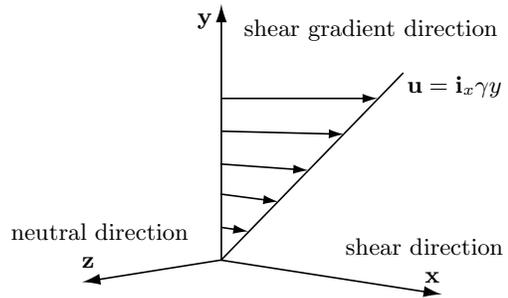}},(32,96)*!LU{\v{y}},(0,10)*!LD{\v{z}},(95,5)*!LD{\v{x}},(5,25)*!CU\txt{neutral direction},(45,94)*!LU\txt{shear gradient direction},(95,20)*!CU\txt{shear direction},(90,75)*!LU{\v{u} = \v{i}_x \gamma y},\end{xy}
\caption{\label{fig:geom} Geometry and kinematics of the planar shear flow}
\end{figure}
We model the systems under conditions of steady planar shear flow (viscometric flow)\cite{UV}. The geometry and kinetics of the flow are illustrated in Fig.~\ref{fig:geom}. The phase space is sampled by integrating the thermostatted SLLOD equations of motion \cite{EM90} for each particle $i\in\left[1, nN\right]$:
\begin{align}
&\dot{\v{q}}_{i} = \v{p}_{i}/m + \v{i}_x \gamma q_{yi}, \\
&\dot{\v{p}}_{i} = \v{F}_{i}   - \v{i}_x \gamma p_{yi} - \lambda \v{p}_{i}, \\
&\dot{d}_x\left(t\right) = \gamma,
\end{align}
where the position of the particle $i$ is $\v{q}_{i}(q_{xi}, q_{yi}, q_{zi})^{T}$, the momentum is $\v{p}_{i}(p_{xi}, p_{yi}, p_{zi})^{T}$, $\v{F}_{i}$ is the force acting on the $i^{th}$ particle, and $d_x$ is the lattice strain associated with the Lees–Edwards periodic boundary conditions. The parameter $\lambda$ is the Gaussian thermostat multiplier:
\begin{equation}
\lambda = \frac{1}{2mK_{0}} \sum^{nN}_{i=1} \left[ \v{F}_{i}\cdot\v{p}_{i} - \gamma p_{xi} p_{yi} \right]
\end{equation}
The Gaussian thermostat fixes the instantaneous kinetic energy along a trajectory, $K(t)=1/{2m} \sum^{nN}_{i=1} \v{p}_{i} \cdot \v{p}_{i} = K(0) \equiv K_{0}$. If the popular Gear predictor-corrector method\cite{Gear, EM84} is used to numerically integrate such equations of motion, the kinetic energy tends to drift away from its initial value. Therefore, a newer symplectic-like integrator obtained in Ref.~[\citeonline{Zhang99}] using an operator-splitting technique is employed, with the integration time step as large as 0.01, which is one order of magnitude greater than in Gear method.

\subsection{Technical details}
Simulations were performed in a cubic cell of size $L = 40$ with periodic boundary conditions for systems of symmetrical diblock copolymers with a fixed chain length of $N = 10+10$. Systems with different concentrations in the range $\rho\sigma^{3}=0.2 \sim 0.8$ were simulated to obtain a zero-shear phase diagram. The effect of shear flow on systems with two concentrations was considered: $\rho\sigma^{3} \equiv n N / L^{3} = 0.3$ with the number of chains $n = 960$, and $\rho\sigma^{3} = 0.4$ with $n = 1280$. Without flow, the first block copolymer system corresponds to a BCC lattice of spherical micelles (S), and the second to hexagonally packed cylinders (C). 
For each shear rate, simulations were carried out from a random configuration until the time dependencies of energy and other variables of interest appeared to be constant. The typical times $t_{0}$ required to reach a steady state varied from as long as $t_{0} = 6\times10^{4}$ at low shear rates to $t_{0} = 4\times10^{3}$ at the highest shear rate $\gamma = 2$. 
 
Solvent molecules surrounding block copolymers are not simulated explicitly in our model. Instead, the main phenomenological effects of the medium such as shear impulse transfer and thermostating are included in the equations of motion, and selectivity is ensured by the effective potentials. Such a mean-field approach to the medium corresponds to the situation when the time and space scales of processes in the solvent are much smaller than those of the solute. Thus, the values calculated represent the corresponding polymer contributions.

\subsection{Static structure factor}
The quantity that helps to examine the structure is the static structure factor, which gives the position and intensity of the peacks in small-angle scattering experiments:\cite{Chaikin}
\begin{equation}
S_{B}\left(\v{q}\right) =  \left\langle \frac{1}{N_{S}} {\left| \sum^{N_{S}}_{j=1} e^{i\v{q} \v{r}_{j}} \right|}^{2} \right\rangle,
\end{equation}
where the summation is performed over $N_{S}$ B-beads that are inside the circle inscribed in the simulation cell. Thus, $S_{B}$ corresponds to a scattering experiment that probes only the order of the B-rich domain structure. The angular brackets denote an ensemble average that also embraces the non-constant $N_{S}$. For the anisotropic systems studied, the scattering patterns depend naturally on the direction of the beam. By choosing the scattering wave vector $\v{q}$ appropriately, the patterns on the three orthogonal planes, with normal along the shear, shear gradient, and neutral directions, were obtained. 


\section{Results}
\subsection{Zero-shear phases}
%
%
\renewcommand{\tabularxcolumn}[1]{>{\small}m{#1}}
\begin{table}
\caption{\label{tab:zeroshear}Phases of symmetrical diblock copolymers in a selective solvent. DS - disordered spherical micelles; S - BCC lattice of spherical micelles; C - hexagonally arranged cylinders; G - bicontinuous, ``gyroid", phase; L - lamellar phase.}
\newcommand{\basiccellwidth}{\hsize}
\newcommand{\smallcellwidth}{0.5\hsize}
\newcommand{\leftcellwidth}{1.5\hsize}
\newcolumntype{B}{@{}>{\small\setlength{\hsize}{\basiccellwidth}\centering\arraybackslash}X@{}}
\newcolumntype{L}{@{}>{\small\setlength{\hsize}{\leftcellwidth}\raggedright\arraybackslash}X@{}}
\newcolumntype{S}{@{}>{\small\setlength{\hsize}{\smallcellwidth}\centering\arraybackslash}X@{}}
\begin{tabularx}{\linewidth}{LSBBBBB}
\hline
\hline
concentration,$\%$ & & 20 & 40 & 55 & 65 & 100\\
\end{tabularx}
\begin{tabularx}{\linewidth}{LB|B|B|B|B|S}
\hline
phase & DS & S & C & G & L & \\
\hline
\hline
\end{tabularx}
\end{table}
%
%
The morphological transitions caused by varying the concentration of symmetrical diblock copolymers in selective solvent in the absence of shear flow are schematically given in Table~\ref{tab:zeroshear}. The diagram qualitatively agrees with the previously published experimental\cite{L02}, theoretical\cite{L98} and lattice Monte Carlo computer simulation\cite{Lar96, Windle01} results. With an increase in concentration, the system adopts the following sequence of morphologies: DS $\to$ S $\to$ C $\to$ G $\to$ L. Therefore, the model and method used in this work can be concluded to be suitable for studying structural changes in block copolymers. However, some expertise and special techniques, e.g. simulating annealing when the temperature of the system being equilibrated is gradually reduced, are required to attain equilibrium for the systems without shear flow. These difficulties can be attributed to the large number of particles (up to as many as $51200$), the associative character of interactions, and the effects of the periodic boundary conditions. In contrast, as will be described later, a steady state is attained much faster and easier under shear flow.

\begin{figure*}\centering
\fourfigs{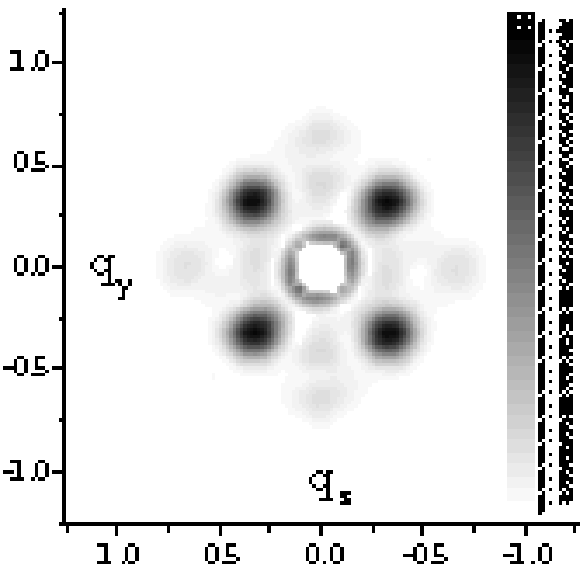}{0}
\fourfigs{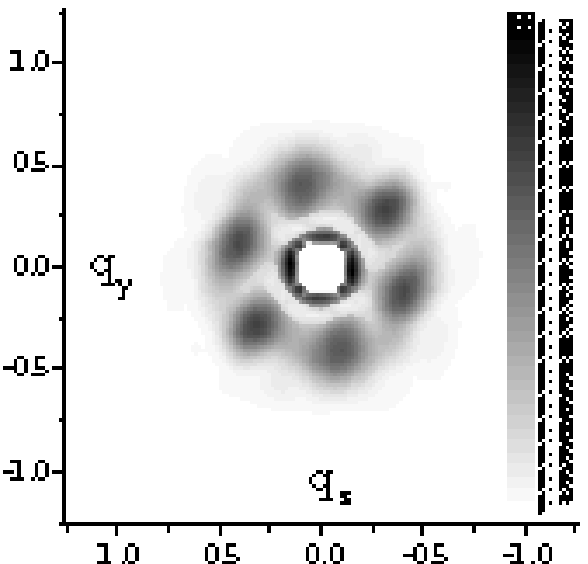}{0.005}
\fourfigs{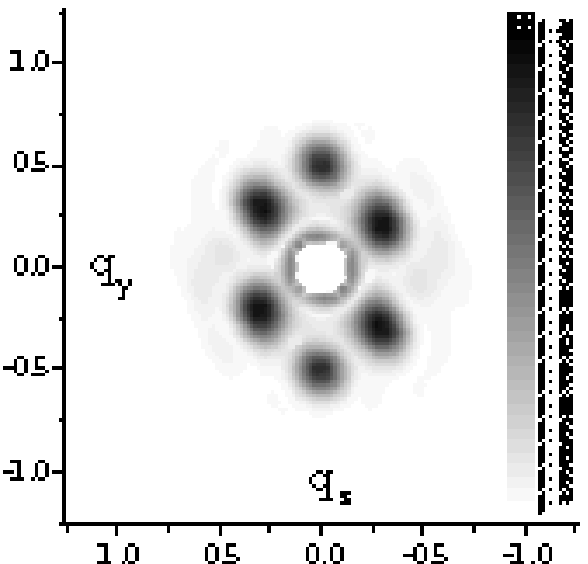}{0.02}
\fourfigs{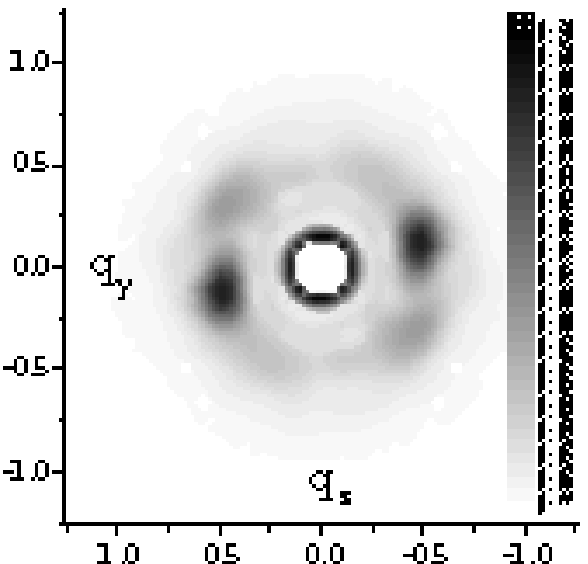}{0.2}
\fourfigs{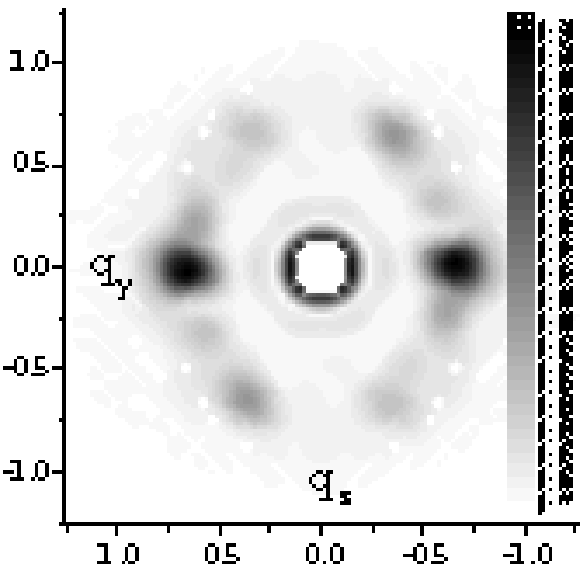}{0.7}
\fourfigs{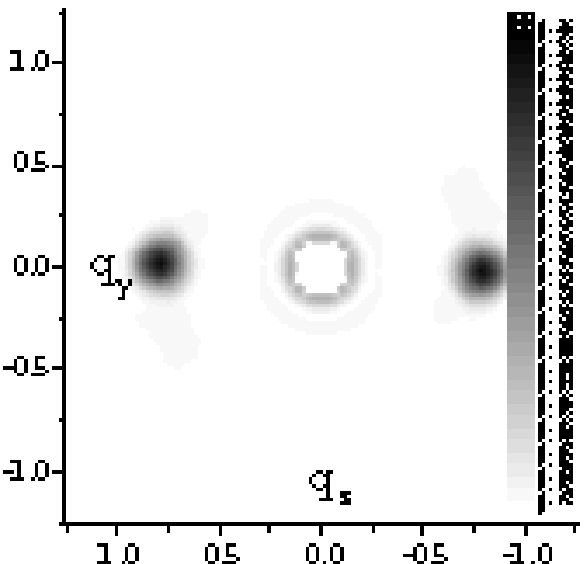}{1.5}
\caption{\label{fig:rho03} Typical snapshots and scattering patterns for the system $\rho\sigma^{3}=0.3$.} 
\end{figure*}
\begin{figure*}\centering
\fourfigs{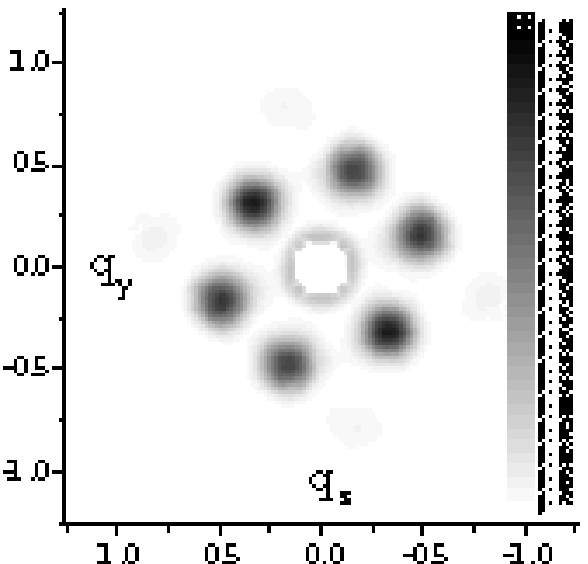}{0.001}
\fourfigs{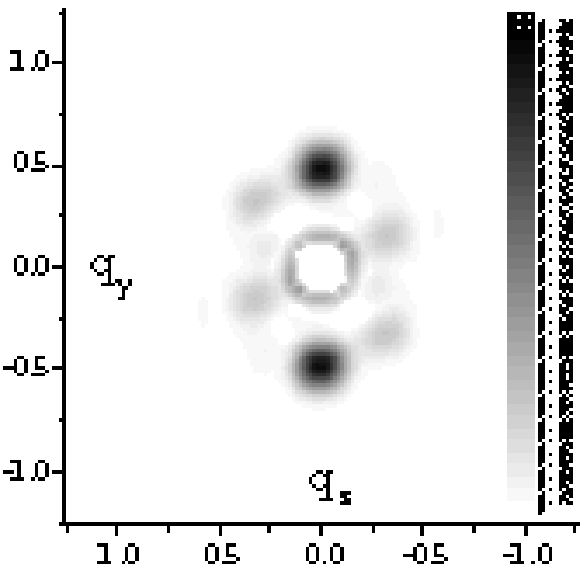}{0.04} 
\fourfigs{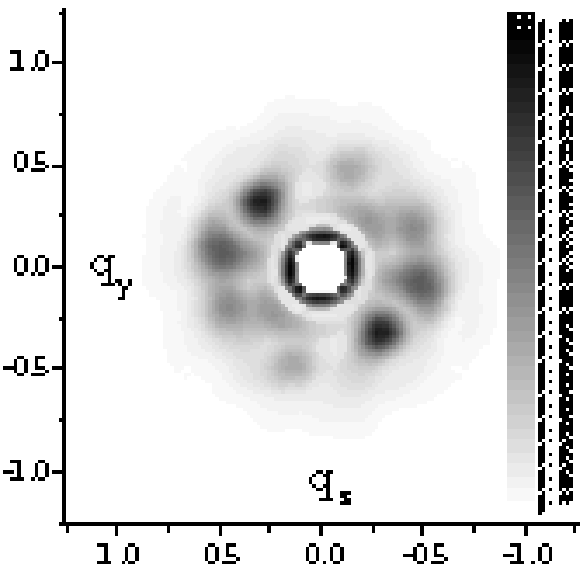}{0.07} 
\fourfigs{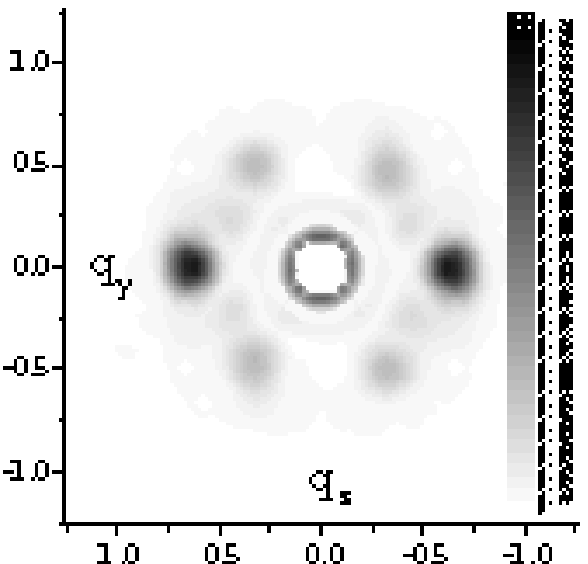}{0.4} 
\fourfigs{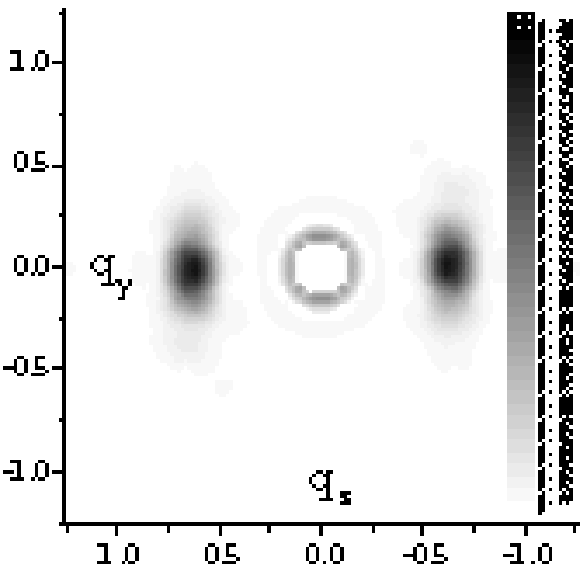}{0.7} 
\fourfigs{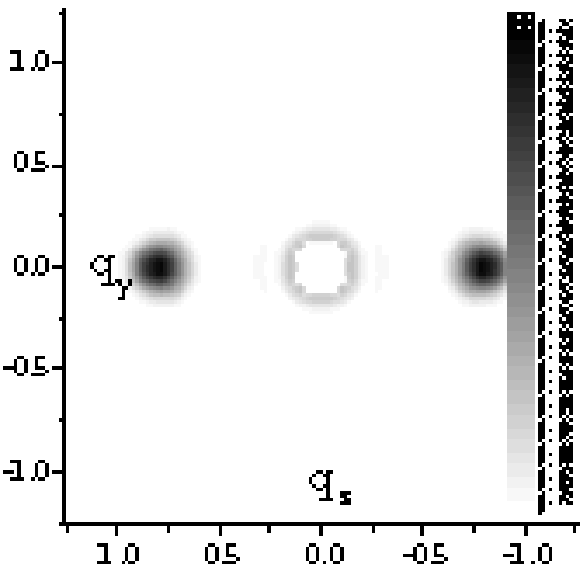}{1.5} 
\caption{\label{fig:rho04} Typical snapshots and scattering patterns for the system $\rho\sigma^{3}=0.4$.} 
\end{figure*}
\subsection{Effect of shear flow}
The structural changes caused by shear flow are illustrated in Figs.~\ref{fig:rho03} and~\ref{fig:rho04}. Each row in both figures contains a typical real-space snapshot of the configuration of B-rich domains, with A-rich domains in the interspacings omitted for clarity. The next three subfigures in a row are gray-scale maps of the  intensity of the structure factor (scattering patterns) projected on the $(zy)$ (shear direction), $(xz)$ (shear gradient) and $(xy)$ (neutral direction) planes. Different rows correspond to the steady state structures observed for different shear rates. They show the evolution, from the top to bottom, of the systems with increasing shear rates.
\subsubsection{BCC micellar crystal sheared}
The results for a polymer density of $0.3$ are presented in Fig.~\ref{fig:rho03}. At  zero shear rate the system exhibits a BCC lattice of spherical micelles. With the  application of shear, the simple cubic symmetry is broken and six-fold patterns appear on the $zy$ pattern \subref{subfig:ro03shear0005}. Such a six-fold pattern has been observed experimentally\cite{Gast95} and was explained as local distortion of the (110) BCC plane into 2-d HCP layers. The layers also reorient, with their normal perpendicular to the shear gradient \subref{subfig:ro03shear002}. Thus, the flow proceeds via parallel sliding of the closely packed layers. With an increase in the shear rate above $\gamma^{*} \approx 0.05$, the sliding parallel layer structure becomes unstable and the phase transition commences \subref{subfig:ro03shear02}, leading to the formation of  perpendicular lamellae. The new structure can be distinguished unambiguously at $ \gamma^{**} \approx 0.4$, and further changes include stabilization and better separation within lammelae \subref{subfig:ro03shear07} and a shear-induced decrease in lamellar spacing \subref{subfig:ro03shear15}. 

\subsubsection{HEX cylinders sheared}

The results for a density of $0.4$ are presented in Fiq.~\ref{fig:rho04}. While it is difficult to achieve a cylindrical phase in HEX ordering in the absence of shear flow, even a small shear leads to a much faster formation of HEX cylinders aligned along the shear direction, \subref{subfig:ro04shear0001}. The structure remains stable after  shear flow is turned off and differs from that obtained without any shear only in terms of the orientation of cylinders. The energy of the state is also lower than that of any apparently metastable state generated without pre-shearing. This suggests that such pre-shearing techniques can be  used to facilitate attaining equilibrium structures in complex fluids like microsegregated block copolymer systems.

Greater shear breaks the percolating cylinders into rod-like micelles. An increase in the shear rate up to the first critical value $\gamma^{*}$ causes the system to rearrange into sliding parallel layers composed of the rod-like micelles \subref{subfig:ro04shear004}.
An increase in the shear rate above $\gamma^{*}$ causes the onset of the transition \subref{subfig:ro04shear007} to perpendicular lamellae, which is completed by the second critical shear rate $\gamma^{**}$ \subref{subfig:ro04shear04}. A stronger shear leads to  enhancement of the normal density fluctuations of lamellae and their splitting, or to  shear-induced decrease in lamellar spacing, which is apparent in the snapshot of the simulation cell as the splitting of four into five lamellae. 

\subsection{Chain conformations}
\begin{figure}\centering
\leavevmode\begin{xy}\xyimport(102,100){\includegraphics[width=7cm]{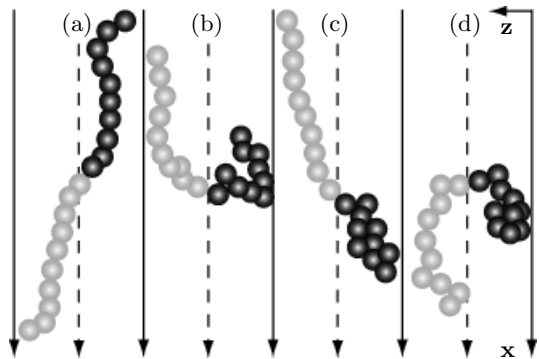}},(10,98)*!LU\txt{(a)},(35,98)*!LU\txt{(b)},(60,98)*!LU\txt{(c)},(85,98)*!LU\txt{(d)},(95,95)*!LU{\v{z}},(95,0)*!LD{\v{x}}\end{xy}
\caption{\label{fig:conf} Typical snapshots of chain conformations in a perpendicular lamellar structure. The chains lie in the $(xz)$ plane, and therefore it is sufficient to only show this projection.}
\end{figure}
Computer simulation of polymeric systems makes it possible to directly observe chain conformations. In the systems studied, due to the selective solvent, the conformations of blocks of $A$ and $B$ species differ. In the absence of shear flow, $A$-blocks are in swollen coil-like conformations while $B$-blocks are in a rather compact globule-like state. Shear flow is known to stimulates polymer chains to elongate and orient in the shear direction\cite{PolFlow}. We observed such a behavior for both $A$-blocks and $B$-blocks, with flow overcoming the associative character of the interaction of the latter. In view of the formation of perpendicular lamellae under high flow, where block copolymer conformations have to be compatible with lamellae spacing in $z$ direction, a question arises of whether the chain orientation is still along the direction of shear. Typical chain conformations in the perpendicular lamellar phase are shown in Fig.~\ref{fig:conf} in the $(xz)$ plane. This projection shows essentially all of the information needed because the chains on average lie in the shear plane. The figure shows that both subchains flip between two directions \emph{on average} oriented along the shear direction. This observation is supported by studies of the subchain gyration tensor which will be published elsewhere.
\section{Conclusions}
%
%
\newcommand{\ics}{0pt} 
\renewcommand{\tabularxcolumn}[1]{>{\small}p{#1}}
\begin{table}
\caption{\label{tab:trends}Structural behaviour of diblock copolymers in selective solvents under shear flow depending on the shear rate.}
\newcommand{\basiccellwidth}{\hsize}
\newcommand{\smallcellwidth}{0.5\hsize}
\newcommand{\leftcellwidth}{0.5\hsize}
\newcolumntype{B}{@{\extracolsep{\ics}}>{\small\setlength{\hsize}{\basiccellwidth}\centering\arraybackslash}X@{}}
\newcolumntype{L}{@{\extracolsep{\ics}}>{\small\setlength{\hsize}{\leftcellwidth}\raggedright\arraybackslash}X@{}}
\newcolumntype{S}{@{\extracolsep{\ics}}>{\small\setlength{\hsize}{\smallcellwidth}\centering\arraybackslash}X@{}}
\begin{tabularx}{\linewidth}{LBBBS}
\hline
\hline
\mbox{shear rate} &  0 & 0.05 & 0.4&\\
\end{tabularx}
\begin{tabularx}{\linewidth}{LS|B|B|B}
\hline
\mbox{BCC micelles} or \mbox{HEX cylinders} & & domain rearranging into parallel layers & domain melting and transition to perpendicular lamellae & perfect perpendicular lamellae; reduction in lamellar spacing  \\
\hline
\hline
\end{tabularx}
\end{table}
%
%

The non-equilibrium molecular dynamics method was applied to study block copolymer systems in a selective solvent under a shear flow field. The systems undergo structural changes that were studied with regard to the patterns of the asymmetric static structure factor and typical 3-dimensional snapshots. The structure factor, unlike the instantaneous snapshots, provides a more reliable basis for drawing conclusions, and showing the patterns in all three mutually perpendicular projections makes it possible to classify the new morphologies. The results enable us to clearly distinguish two trends, as shown in Table~\ref{tab:trends}. In weak flow, the systems show parallel sliding of close-packed layers, with the normal along the shear gradient. At higher shear rates, the more favorable morphology is perpendicular lamellae, with the normal along the neutral direction. This behavior is observed for both systems with two different polymer concentrations.

The following explanation can be adopted from studies of block copolymer melts, where the parallel and perpendicular orientation of the lamellar phase under weak and strong shear flow, respectively, has been investigated\cite{F94, F96}. In the parallel alignment, less viscous strata permit the easy sliding of mechanically contrasting layers, which increases the basin of attraction of this steady-state orientation. However, near the ODT, where the mechanical contrast is not so large, stronger shear leads to perpendicular orientation. It has been argued that a steady flow suppresses concentration fluctuations and a weak shear more strongly suppresses fluctuations along the shear gradient, thus stabilizing parallel orientation. While a stronger shear more strongly suppresses fluctuations along the vorticity axis.

In our study, we observed that flow via parallel and perpendicularly oriented layers or lamellae is also seen for block copolymer solutions.

Two critical values of the shear rate are found. One, $\gamma^{*} \approx 0.05$, corresponds to the end of the region of the stability of parallel sliding layers and the onset of the transition leading to perpendicular lamellae, which are clearly formed at $\gamma^{**} \approx 0.4$. For even higher shear rates, $\gamma > \gamma^{**}$, shear  merely reduces the lamellar spacing. The significance of these critical values can be discussed in relation to the shear rate-dependence of the rheological and microscopic properties, which will be discussed elsewhere.

The parallel sliding layers result from rearrangement of the domains of the original zero-rate structures, whereas the formation of perpendicular lamellae is accompanied by  significant breaking-up and remelting of the domains. This observation should provide a basis for theoretical treatment and provide a subject for further experimental study: under shear flow strong enough to overcome the segregation tendencies, do block copolymers in a selective solvent adopt a perpendicular lamellar structure?

\ifrevtex
\bibliography{../myabbrev,../all}
\else

\fi

\end{document}